# Boundary-Preserved Deep Denoising of the Stochastic Resonance Enhanced Multiphoton Images

**SHENG-YONG NIU[1,2], LUN-ZHANG GUO[3], YUE LI[4], TZUNG-DAU WANG[5], YU TSAO[1,*], TZU-MING LIU[4,*]**

1. *Research Center for Information Technology Innovation (CITI), Academia Sinica, Taipei, Taiwan.*
2. *Department of Computer Science and Engineering, University of California San Diego, CA, USA.*
3. *Institute of Biomedical Engineering, National Taiwan University, Taipei 10617, Taiwan.*
4. *Institute of Translational Medicine, Faculty of Health Sciences, University of Macau, Macao SAR, China.*
5. *Cardiovascular Center and Division of Cardiology, Department of Internal Medicine, National Taiwan University Hospital and College of Medicine, Taipei 10002, Taiwan.*

** yu.tsao@citi.sinica.edu.tw and tmliu@umac.mo*

**Abstract:** As the rapid growth of high-speed and deep-tissue imaging in biomedical research, it is urgent to find a robust and effective denoising method to retain morphological features for further texture analysis and segmentation. Conventional denoising filters and models can easily suppress perturbative noises in high contrast images. However, for low photon budget multi-photon images, high detector gain will not only boost signals, but also bring huge background noises. In such stochastic resonance regime of imaging, sub-threshold signals may be detectable with the help of noises. Therefore, a denoising filter that can smartly remove noises without sacrificing the important cellular features such as cell boundaries is highly desired. In this paper, we propose a convolutional neural network based denoising autoencoder method, Fully Convolutional Deep Denoising Autoencoder (DDAE), to improve the quality of Three-Photon Fluorescence (3PF) and Third Harmonic Generation (THG) microscopy images. The average of the acquired 200 images of a given location served as the low-noise answer for DDAE training. Compared with other widely used denoising methods, our DDAE model shows better signal-to-noise ratio (26.6 and 29.9 for 3PF and THG, respectively), structure similarity (0.86 and 0.87 for 3PF and THG, respectively), and preservation of nuclear or cellular boundaries.

## 1. INTRODUCTION

The real-time visualization of living cells in their tissue environment is crucial for many applications in life science and medical devices [1, 2]. For instance, deep-tissue cellular imaging with transgenic labeling of reporters could reveal the niche environments and functional interactions of multiple cells in the contexts of hematopoiesis [3], tumor metastasis [4-6], and neuronal connection [7, 8]. *In vitro* or non-invasive *in vivo* imaging flow cytometry through real-time microscopy of the intra-fluidic channel or intravascular blood cells could have the potential to benefit patients and caregivers to detect physiological aberrancy more efficiently [9-13]. To obtain sharp sectioning images of cells in tissues, researchers developed confocal or multiphoton fluorescence microscopy to improve the axial resolution [1, 14]. Two-photon fluorescence microscopy, for instance, has been widely applied in many deep-tissue studies such as brain research. Its near infrared (800-1300 nm) laser excitation can greatly reduce the photo-bleaching of probes, distortion of wave-front, a scattering of photons, and maintain the subcellular resolution of images [1, 15, 16]. For good enough contrast of fluorescence imaging at deep tissues, the excitation intensity needs to be high, which might lead to background interference originated from the diffused fluorescence photons caused by multiple scattering. To obtain high acuity images at a depth deeper than 700 μm, three-photon contrasts excited by 1700 nm high pulse energy laser. The three-photon microscopy can improve significantly in the emission localization, reducing the out-of-focus background compared to two-photon microscopy [17]. Various existing fluorescent dyes can still work in 3PF microscopy. On the other hand, to realize *in vivo* imaging and medical testing such as virtual optical biopsy of immune cells, *in vivo* label-free microscopy would be critical. Many researchers have put efforts in the development of THG Microscopy for label-free imaging in tissues [11, 18-21]. At an excitation wavelength of 1230 nm, THG microscopy can non-invasively obtain cellular morphology and visualize subcellular organelles in deep tissues without labeling. It can deliver an alternative contrast modality to complement multi-photon fluorescence microscopy, which provides information on cellular morphologies in three-dimensional tissue culture [16, 22].

Although there have been several studies in both 3PF and THG microscopies, their low signal-to-noise ratio is still a crucial issue for the delineation and segmentation of cells. The signal counts of third-order nonlinear optical microscopy are pretty low and comparable to the noise counts resulted from the high bias-voltages of detection units. Type of noises include the signal dependent Poisson noises and detector dependent Gaussian noises. The former involves the random process of photon emission and the discrete nature of photoexcited charges. The latter typically results from the flicker noise or thermal noise in electronic systems. At such low photon budget condition, the signals may have stochastic resonance effects, where sub-threshold signals could be boosted over the threshold with the help of detector noises [23]. The signal pattern within cells will carry features of noises, and it's difficult to extract the true signals out from the background. Hence, it is crucial to find an effective noise filtering method to enhance image contrast while retaining the structural information for further segmentation and texture analysis. Typically, researchers applied the Gaussian filter and median filter to remove Poisson or Gaussian noises. For Poisson noise, it can be transformed by stabilizing method such as Anscombe transformation into Gaussian white noise, and utilize Gaussian noise filter to alleviate it [24]. The non-local mean method performs noise filtering on image patches with similar textures. Some block-matching and collaborative filtering methods such as Block-matching and 3D filtering (BM3D) have been proposed and widely used [25]. Also, some researchers use Bayesian method trying to extract noise information from the prior knowledge of images, train the likelihood function to predict the residual images V, and subtracted them from noisy observation [24, 26]. These methods can perform well in many image de-noising problems. However, conventional signal processing assumes that noise is a linear addition on signals by $y = S+V$, where $S$ represents signals and $V$ represents noise. This model is proper when the noise $V$ is perturbative to $S$. However, in stochastic resonance regime, $V$ is comparable to or larger than $S$ and the detection threshold $T$ is larger than $S$ in many pixels of images. The representation of overall signals should be $y = S + V - T \text{ if } S + V > T; 0 \text{ if } S + V < T$ . Therefore, conventional de-speckle filter, median filter, or Gaussian filter may more or less drop the details of cellular morphology. For the signal-noise entangled stochastic resonance images, these methods may not work well due to the difference of noise modeling, i.e. the noisy observation is not just a superposition of signals and noise. Hence, for low photon-budget multiphoton microscopy images acquired in high-speed or deep-tissue imaging, it's still a challenging task to enhance signal-to-noise ratio without sacrificing structural information.

Several machine learning and deep learning-based denoising methods have been proposed and proved to generate better filtering results comparing with traditional filters. Especially, deep convolutional neural network (CNN) based algorithms have been widely applied in image classification [27], segmentation [28], and denoising [29, 30]. Recently, based on residual learning and batch normalization in Res U-Net, the content aware image restoration (CARE) method demonstrated good performance in denoising and resolution improvement of low photon-budget microscopy images [30]. Besides, in acoustics area, researchers proposed CNN-based methods and denoising autoencoder (DAE) architectures for noise reduction [31, 32]. The DAE method successfully filtered out background noise and improve the perceptual evaluation of speech quality (PESQ). Inspired by the success of CNN and DAE methods, in this work, we proposed a Fully Convolutional Deep Denoising Autoencoder (DDAE) method to reduce the noise in low photon-budget multiphoton microscopic images, especially for three-photon fluorescence and third-harmonic generation images. Without residual learning or batch normalization structures, the DDAE model has less limitation on its hardware implementation in the future. After training with 28 sets of microscopy data, we found the compact DDAE model already outperforms Gaussian filter, median filter, and benchmark BM3D algorithm on signal-to-noise ratio and structure similarity. Moreover, DDAE well preserves stochastic resonance enhanced features so that regions of nuclei or cells can be delineated more correctly, which is important for further segmentation works.

## 2. METHODS

### *2.1 Cell Culture, Cell Staining, and Acquisition of Multiphoton Images*

RAW 264.7 murine macrophage cell lines were plated on bottom glass dishes (Nest Scientific, 801001), and were cultured in Dulbecco's Modified Eagle's Medium (DMEM) containing 10% fetal bovine serum (FBS), 100 U/ml penicillin and 100 µg/ml streptomycin. For 3PF imaging, no further treatment was added. For THG imaging, three hours after plating, 50 ng/ml lipopolysaccharides (LPS; Sigma) were used to elicit inflammatory macrophages at M1 state. After 24 hours of cytokine stimulation, media was replaced with normal media for cell imaging. For the 3PF imaging, 2 µg/mL Hoechst 33342 (Thermo Fisher Scientific) was used to stain the cell nuclei for 5 min. Unloaded Hoechst 33342 was removed by washing with normal medium. For the THG imaging, there was no labeling on cells. The lipid granules in M1 activated macrophages can give strong THG signals.

Time-lapse 3PF and THG images were acquired on an inverted multi-photon microscope ($A1MP^{+}$, Nikon, Japan). A near-infrared (800-1300 nm) femtosecond laser (InSight X3, Spectra-Physics, Mountain View, California) with 100-fs pulse width and an 80-MHz repetition rate was used as the excitation source. The operation wavelength for 3PF and THG images were 1250 nm, which has the least on-focus phototoxicity and deepest penetration depth for biomedical samples. The laser light first transmitted an 820 nm edged multiphoton dichroic beam splitter and then was focused through a water-immersed 40× and 1.15 NA objective. To avoid the photobleaching, the Hoechst blue labeled cells were excited at an average power of 11 mW (100 GW/cm$^2$ instantaneous intensity), while for non-bleachable THG imaging, 37 mW (335 GW/cm$^2$ instantaneous intensity) was required to obtain detectable signals. All the generated three-photon signals were epi-collected by the same objective, reflected by the 820 nm edged multiphoton dichroic beam splitter, further reflected by a 495 nm edged dichroic beam splitter in the non-descanned detection unit, filtered by a 415-485 nm bandpass filter, and finally detected by the same photomultiplier tubes. Then the laser was raster scanned by a pair of the resonant scanner and a galvanometer mirrors to perform point-by-point excitation and detection, forming 512 × 512 pixels images at a 15-Hz frame rate. All images were subsequently exported to TIFF format images for denoising and deep learning processes.

### *2.2 Traditional Denoising Methods*

For Gaussian and median filter, we used the python scipy function ndimage.gaussian_filter and ndimage.median_filter to perform the Gaussian and median filtering with a standard deviation σ (sigma values) of 1, 3, 5, and 10. For Block-matching and 3D filtering (BM3D) method, we implemented the MATLAB codes from http://www.cs.tut.fi/~foi/GCF-BM3D/ to perform BM3D denoising with a noise standard deviation σ (sigma values) of 120, 140, 160, 180, 200, 220, and 240.

### *2.3 Fully Convolutional Deep Denoising Autoencoder Model*

We used Keras framework to implement denoising autoencoder with a fully convolutional neural network architectures. For 3PF DDAE model, we used 22 set of training 3PF images, 6 set of validation 3PF images and trained the model with 30 epochs. For THG DDAE model, we used 22 set of training THG images, 6 set of validation THG images and trained the model with 20 epochs. Below are the pseudo-codes and architecture of DDAE algorithm.

```
DDAE(input_img):

  // encoder
  conv1 <- Conv2D(32, (3, 3), activation <- 'relu')(input_img)
  pool1 <- MaxPooling2D(pool_size <- (2, 2))(conv1)
  conv2 <-Conv2D(64, (3, 3), activation <- 'relu')(pool1)
  pool2 <- MaxPooling2D(pool_size <- (2, 2))(conv2)
  conv3  <-  Conv2D(128, (3, 3), activation <- 'relu')(pool2)

  // decoder
  conv4  <-  Conv2D(128, (3, 3), activation <- 'relu')(conv3)
  up1  <-  UpSampling2D((2,2))(conv4)
  conv5  <-  Conv2D(64, (3, 3), activation <- 'relu',)(up1)
  up2  <-  UpSampling2D((2,2))(conv5)
  decoded  <-  Conv2D(1, (3, 3), activation <- 'sigmoid')(up2)

  return decoded
```

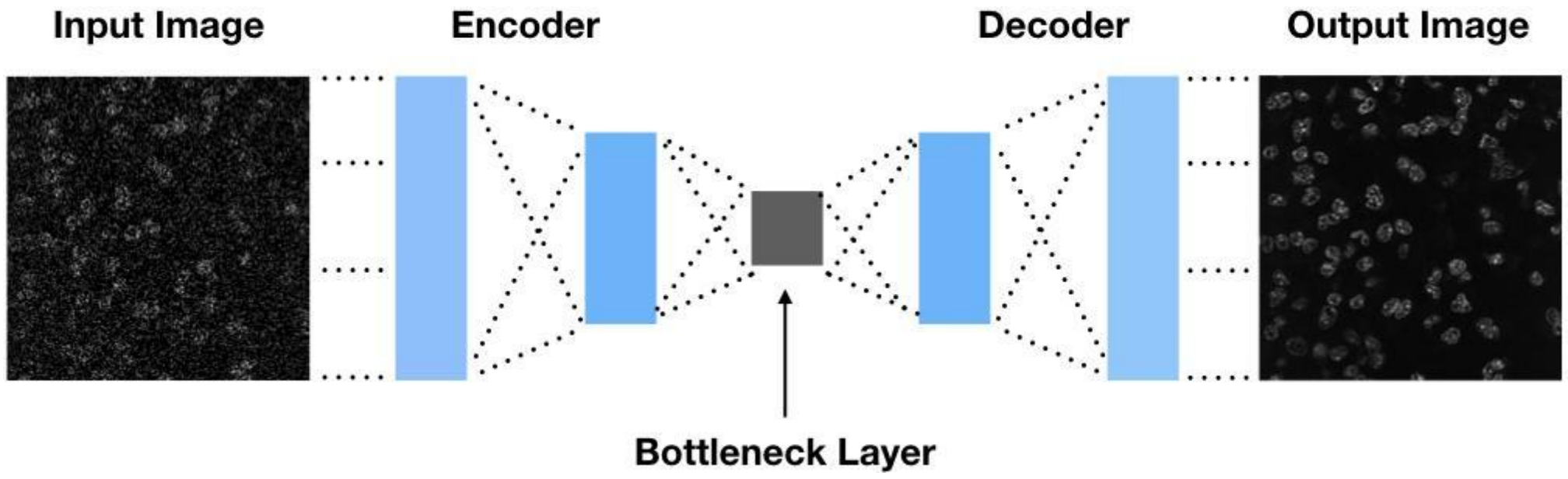

**Fig. 1.** The architecture of Fully Convolutional Deep Denoising Autoencoder model.

### *2.4 Analyzing the Contrast and Structure Similarity of Restored Images*

To evaluate the signal-to-noise ratio and the restoration of structural information, we used the Peak Signal to Noise Ratio (PSNR) analysis [33] and the Structural Similarity Index (SSIM) [34].

### *2.5 Analyzing the Restoration of Nuclear and Cellular Boundaries*

To assess how well the denoising filter retains the stochastic resonance enhanced features in cells, we analyzed the precision, recall, and F-measure of nuclear and cellular boundaries from the denoised image $\mathbf{Img_O}$ and the low-noise answer image $\mathbf{Img_{Ans}}$. The contrasts of $\mathbf{Img_O}$ and $\mathbf{Img_{Ans}}$ were enhanced through histogram equalization first (See Fig. 2), and then binarized by intensity auto-threshold method, IsoData, thus obtaining the binary images of $\mathbf{bImg_O}$ and $\mathbf{bImg_{Ans}}$, respectively. The boundaries of binarized answer images can precisely depict the boundaries of cells and nuclei [Fig. 3(a)]. The $\mathbf{bImg_{AND}}$ was the overlap of $\mathbf{bImg_O}$ and $\mathbf{bImg_{Ans}}$, representing the true positive of nucleus (in 3PF images) or cell (in THG images) regions. Finally, we used the pixel counts of those three binary images, $\mathbf{count_O}$, $\mathbf{count_{Ans}}$, and $\mathbf{count_{AND}}$, to compute the precision, recall and F-measure of the denoised images by the functions:

$$\text{precision} = \frac{\mathbf{count}_{AND}}{\mathbf{count}_{\mathrm{O}}} \qquad (1)$$

$$\text{recall} = \frac{\mathbf{count}_{\mathrm{AND}}}{\mathbf{count}_{\mathrm{Ans}}} \qquad (2)$$

$$\mathrm{F-measure} = \frac{2*\text{precision}*\text{recall}}{\text{precision}+\text{recall}} \qquad (3)$$

Also, we made pseudocolor images to visualize the retained regions of nuclei or cells (blue color, Fig. 2), the false negative regions of $\mathbf{bImg_{Ans}}$ **- bImgAND** (green color, Fig. 2), and false positive regions of $\mathbf{bImg_O}$ **- bImgAND** (magenta color, Fig. 2). Then we combined these three pseudo color images to obtain the superposition image [Fig. 3(b)], such that mismatch of boundaries can be visualized.

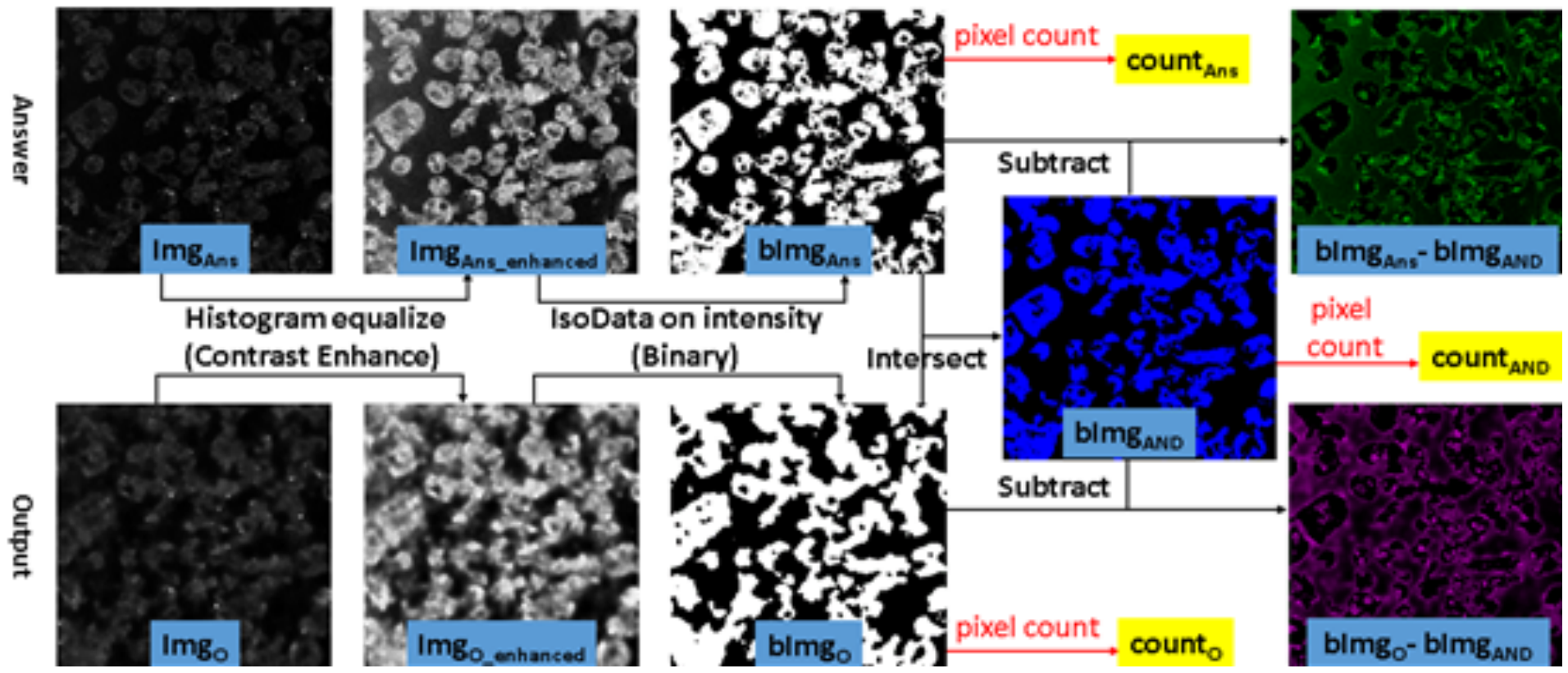

**Fig. 2.** The procedures of Image processing for analyzing the preservation of nuclear and cellular boundaries.

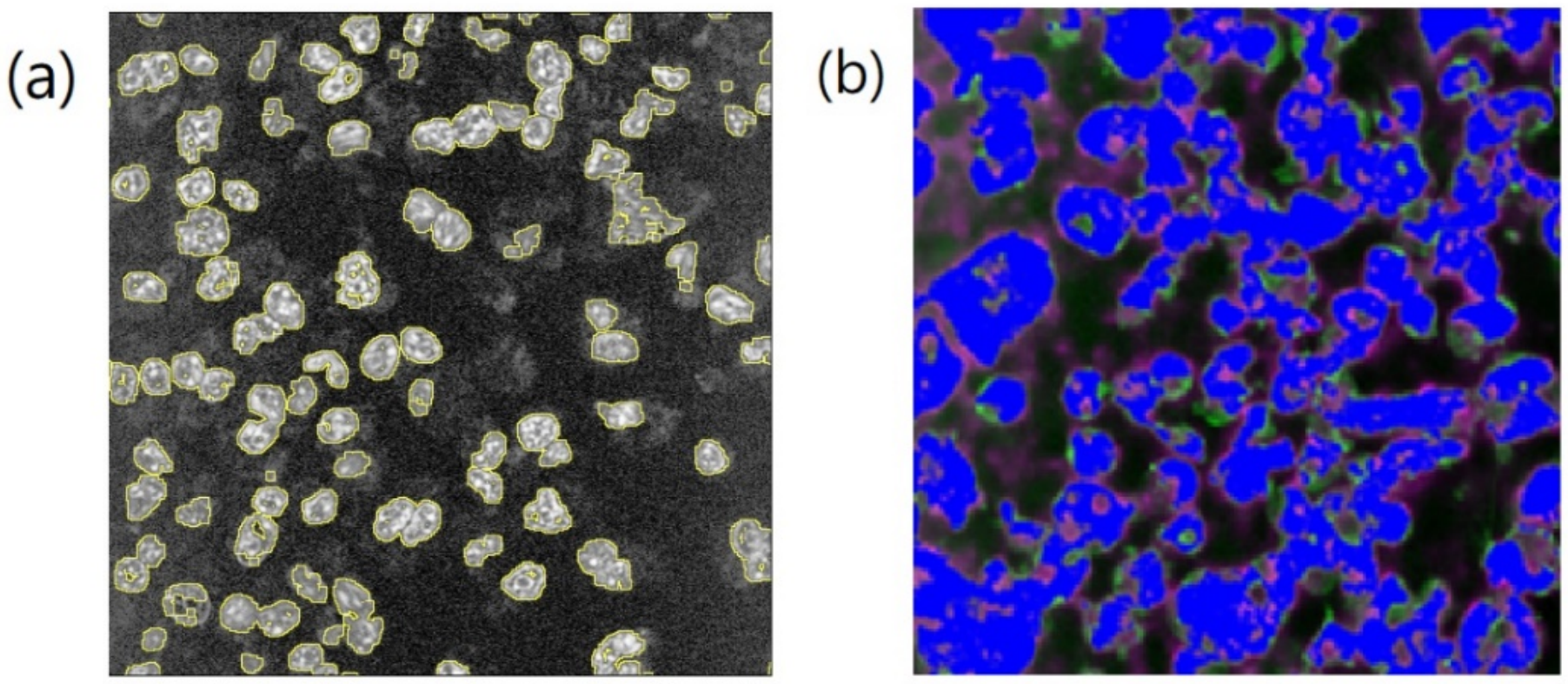

**Fig. 3.** (a) The boundaries (yellow contours) of the binarized 3PF answer image $\mathbf{bImag_{Ans}}$ can precisely outline the nuclear boundaries. (b) The superposition image of the true positive (blue color), the false positive (magenta color), and the false negative (green color) parts of denoised THG images. Fields of view: (a) 120 × 120 µm; (b) 160 × 160 µm.

## 3. RESULTS

For the evaluation of denoising performance among different approaches, we performed 1250 nm excited 3PF and THG microscopy on the RAW 264.7 macrophages. The 3PF contrast majorly labeled the nuclei and the THG contrast revealed the lipid granules within cells. The excitation intensity used for 3PF microscopy is too low to generate sufficient THG signals. Besides, for 3PF imaging, we didn't activate the proliferation of lipid granules in RAW cells. There would be no cross-talks between 3PF and THG signals. At each observation location in a petri dish, we acquired 200 images at a 15 Hz frame rate, with fixed excitation power, by the same detection channel, and using the same bias voltage. They served as the low photon-budget images to be denoised. To evaluate the performance of denoising, the low-noise answer image of the corresponding location was obtained from the average of the acquired 200 images. It also served as the low-noise answer for DDAE training. For each imaging modality, we picked up 31 locations on each petri dish and acquired 31 batches of images. The 28 batches of them were used for DDAE as training and validation sets and the other three batches for testing. Then we compared the results with those processed by traditional denoising methods such as Gaussian filter, median filter, and BM3D. We measured the quality of results by PSNR and SSIM to represent the fidelity of signal and structures.

### *3.1 Denoising of Three-Photon Fluorescence Images*

Three testing cases of 3PF images were sampled from each testing batches, respectively. Since the images were acquired at a 1/15 sec frame time, they contained a lot of salt-and-pepper noises [Fig. 4(a), upper row]. The stochastic resonance enhanced signals made cells faintly discernible [Fig. 4(c)]. Applying the trained DDAE model, they showed a great improvement in image contrast [Fig. 4(a), bottom row]. The nuclear boundaries were well-preserved [Fig. 4(d), yellow dashed contour]. For cells with relatively low signal level in low-noise answer image [Fig. 4(b), pointed by blue arrows], noises can't boost them up in the high frame-rate image, and the DDAE can't restore them in such situation. Compared with traditional filtering methods such as Gaussian filter, median filter and BM3D (Fig. 5), the results of 3PF DDAE denoising surpass all of them on the PSNR and SSIM scores (Table 1 and Supplementary Table 1). In general, the DDAE method achieves a higher signal-to-noise ratio of 26.6 PSNR and retains more structural information of 0.86 SSIM. Among the traditional methods, by choosing optimal sigma value, the performance of the Gaussian filter is the best with 25.59 PSNR. It shows a relatively clear cellular outline but loses many intracellular details with 0.8 SSIM. As for the results in the median filter, although it can retain a few significant signal spots, it cannot show nuclear boundaries. The BM3D method performs better than the median filter in general, but it contains significant noise and vague nuclear boundaries in the results.

(a)

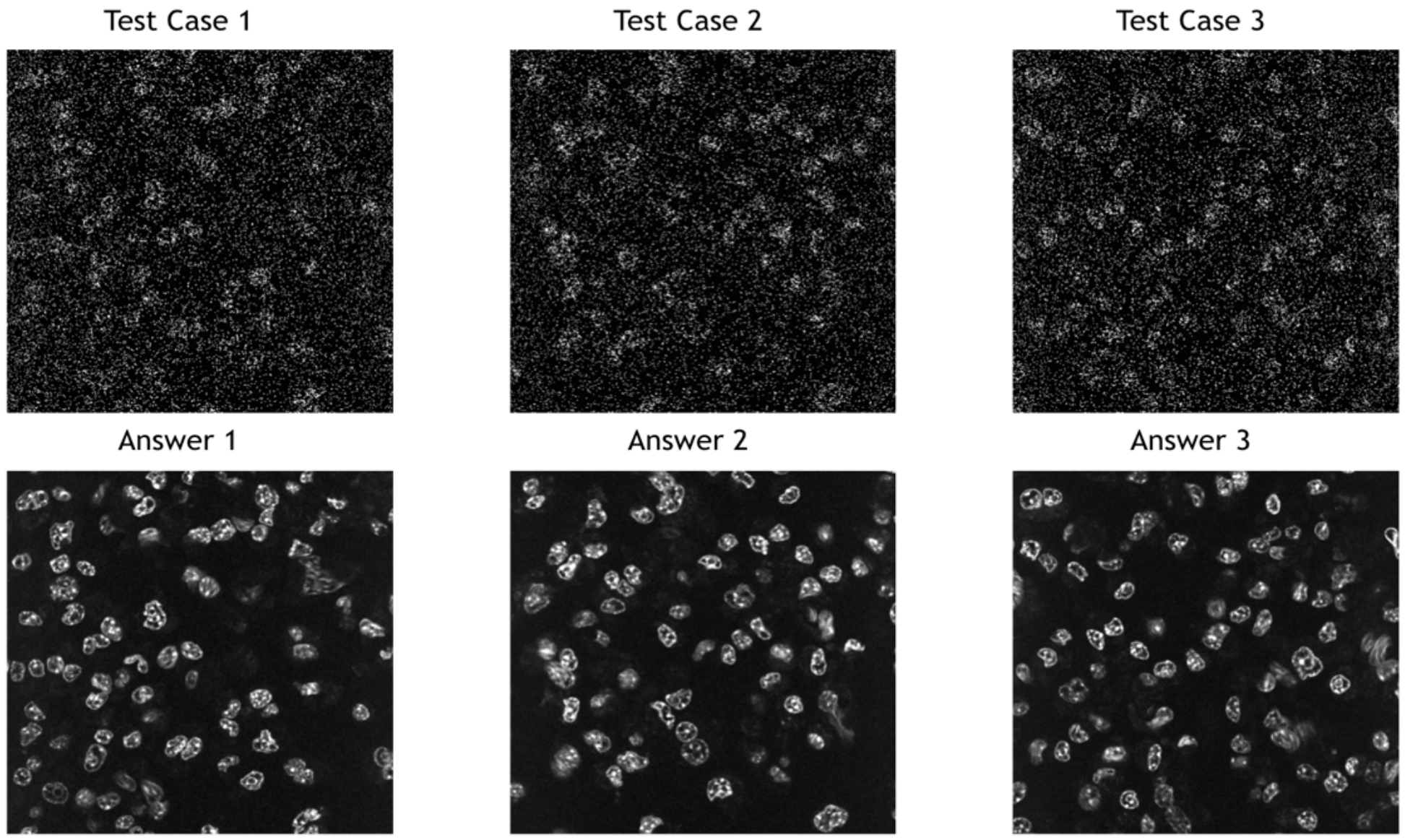

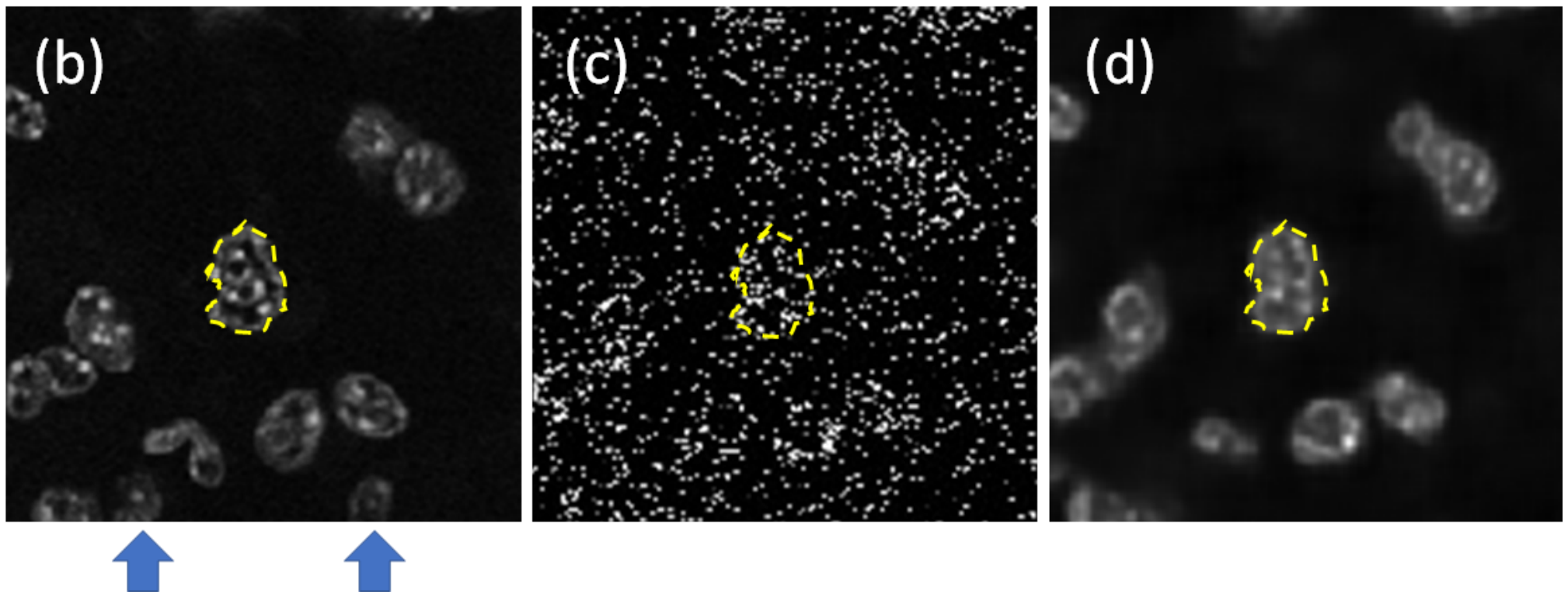

**Fig. 4.** (a) Noisy inputs (upper rows) and low-noise answers (bottom rows) of three testing 3PF images of Hoechst blue labeled RAW cells. (b) The low-noise answer image, (c) noisy input, and (d) DDAE processed one. Processed by DDAE model, the noise was suppressed, the contrast was enhanced, and the nuclear boundary was well-preserved (yellow dashed closure). Fields of views: (a) 120 × 120 µm; (b-d) 50 × 50 µm.

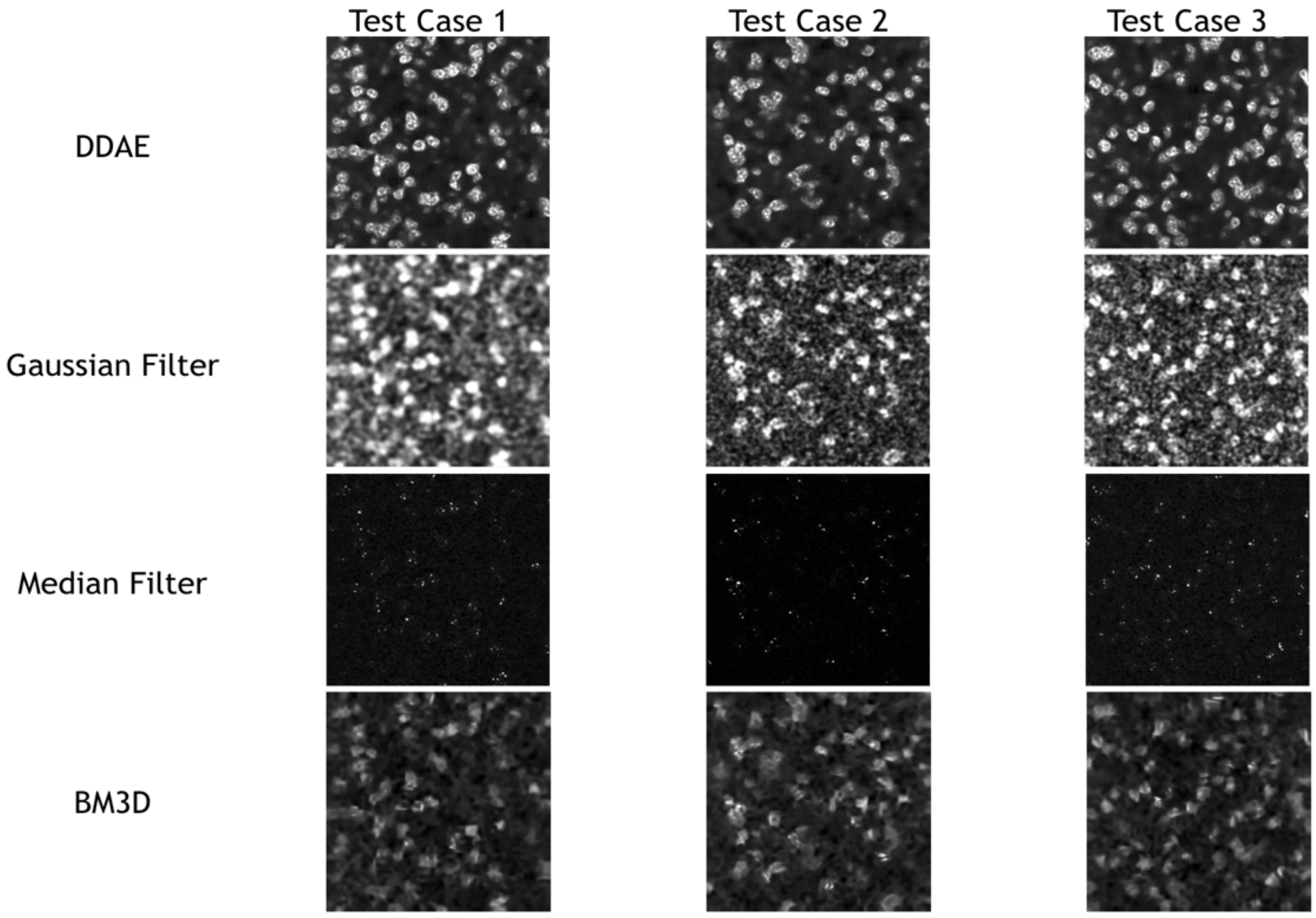

**Fig. 5.** Denoising results of 3PF images with DDAE, Gaussian filter, median filter, and BM3D algorithms.

**Table 1. Average PSNR and SSIM Scores of Denoising Results**

| | 3PF Images | | THG Images | |
|---|---|---|---|---|
| **Filter** | **PSNR** | **SSIM** | **PSNR** | **SSIM** |
| DDAE | 26.60 | 0.86 | 29.92 | 0.87 |
| Gaussian | 25.59 (σ=3) | 0.80 (σ=3) | 29.34 (σ=5) | 0.86 (σ=5) |
| Median | 18.72 (σ=5) | 0.16 (σ=5) | 22.27 (σ=5) | 0.24 (σ=5) |
| BM3D | 25.12 (σ=160) | 0.72 (σ=160) | 29.49 (σ=140) | 0.75 (σ=140) |

In brief, thus trained 3PF DDAE model can improve the image qualities significantly better than all listed traditional denoising methods, and users can have the result in a few seconds without the time-consuming steps like trying optimal sigma values.

### *3.2 Denoising of Third Harmonic Generation Images*

Similarly, we built the THG DDAE model by 28 batches of training and validation sets, composed of 200 THG images acquired at 28 different locations on the petri dish. Instead of nuclei, the THG images reveal the granules in the cytoplasm (Fig. 6), which delineates the outline of cells. Results (Fig. 7) show that the THG DDAE model also outperforms all traditional methods in both PSNR and SSIM scores (Table 1 and Supplementary Table 2). Besides, among the traditional methods, PSNR of BM3D and SSIM of Gaussian filter are the best of them. In the Gaussian filter, we can find that it can retain a general structure of cells, but there is still significant noise in the background. The median filter, similar to the case in 3PF images, is still hard to retain the structures and signals of the cells. As for the BM3D method, it performs relatively well in its signal to noise ratio, but still becomes vague at the cellular boundaries. In general, the THG DDAE model is the best choice for denoising THG microscopy images in the aspects of PSNR (=29.92) and SSIM (=0.87).

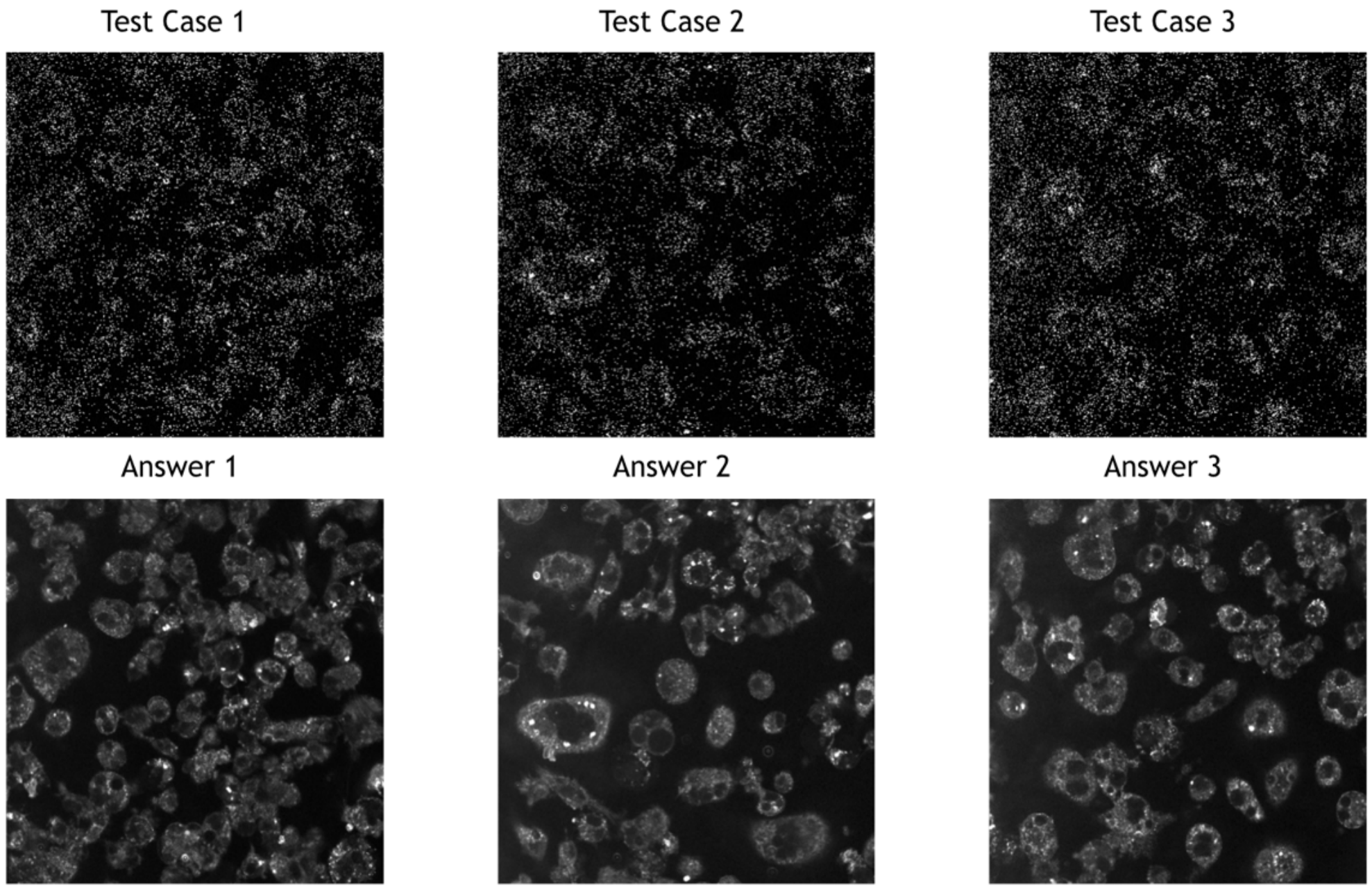

**Fig. 6.** Noisy inputs (upper rows) and low-noise answers (bottom rows) of three testing THG images of RAW cells. Fields of views: 160 × 160 μm.

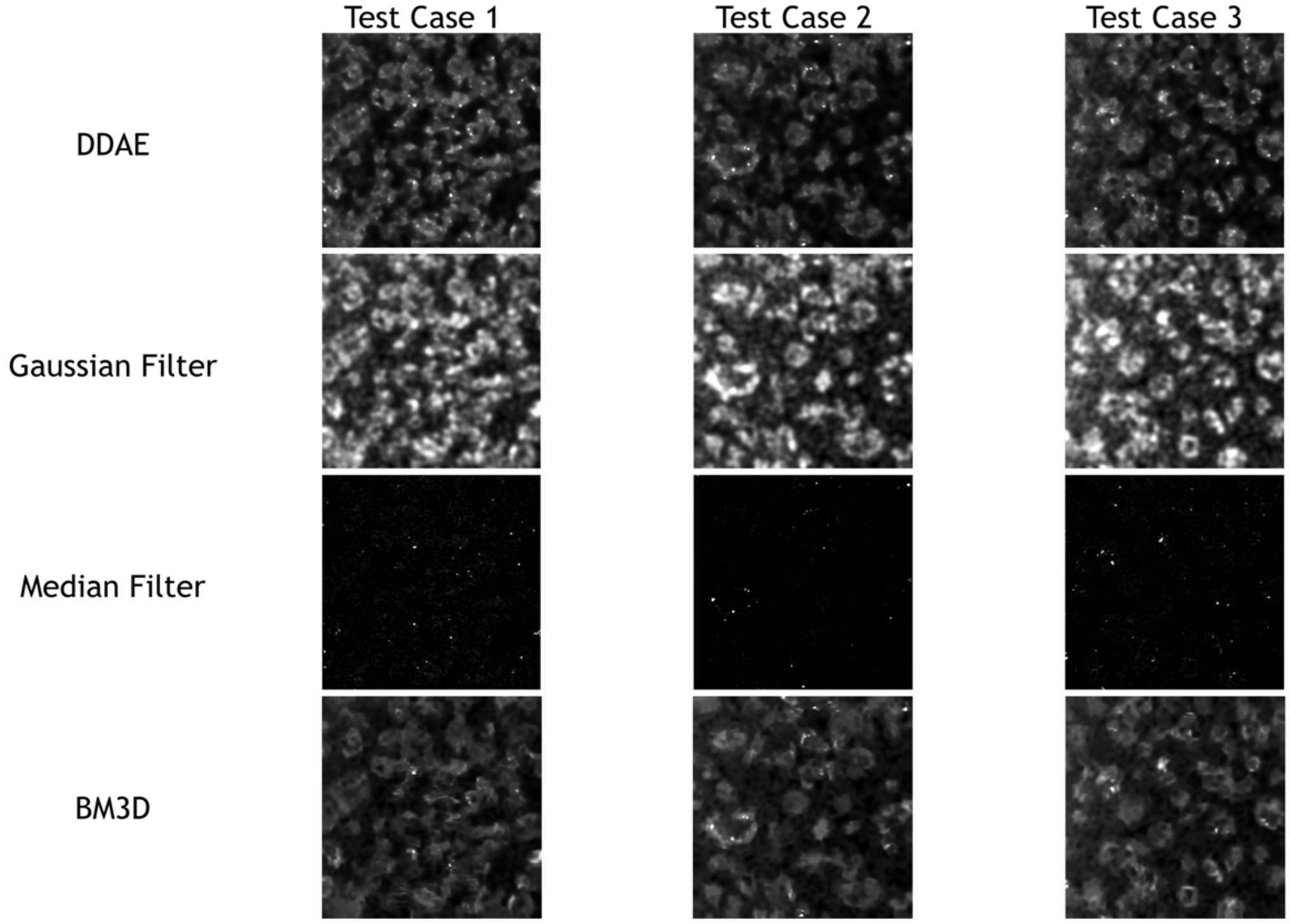

**Fig. 7.** Denoising results of THG images with DDAE, Gaussian filter, median filter, and BM3D algorithms.

### *3.3 The Preservation of Nuclear and Cellular Boundaries*

To understand whether DDAE can preserve more stochastic resonance enhanced signals than other denoising filters do, following the procedures described in the Methods section, we identified the pixels that represent the regions of nuclei or cells in both denoised and low-noise answer images. Then we computed their precision, recall rates and F1 scores (Table 2 and Supplementary Table 3) to evaluate how well the boundary information was preserved after denoising (See Fig. 8, 9). For the 3PF cases, we found that the 70% precision rate of DDAE mode is higher than most of the filters, except for a few cases of median filters ($\sigma$=3 or 5) in extreme situations which have very low recall rates of 10% and 7%. The 77% recall rate of DDAE performed better than most Gaussian and median filters, but not as good as BM3D (86-88%). This indicates that DDAE generates more false negatives (green color in Fig. 8) than BM3D in the analysis of nuclei boundaries. Balanced with the F1-score, which takes precision and recall as equal weighting, the DDAE model has an average 0.734 score much higher than other filters. For the THG cases, the signal level and dynamic range of the low-noise answer image is much lower than 3PF. We found that the average 76% precision rate of DDAE is lower than filters such as BM3D and parts of Gaussian and Median filters' result, whereas the DDAE's average 88% recall rate is the best among all of them. This result indicates that DDAE may result in more false positive pixels and less false negative pixels in THG images of cytoplasm. Balanced with the F1-score, the 0.821 score of DDAE model again outperforms other filters. These results indicated that, for low photon-budget images containing stochastic resonance enhanced signals, DDAE can retain nuclear or cellular boundaries more correctly for further segmentation.

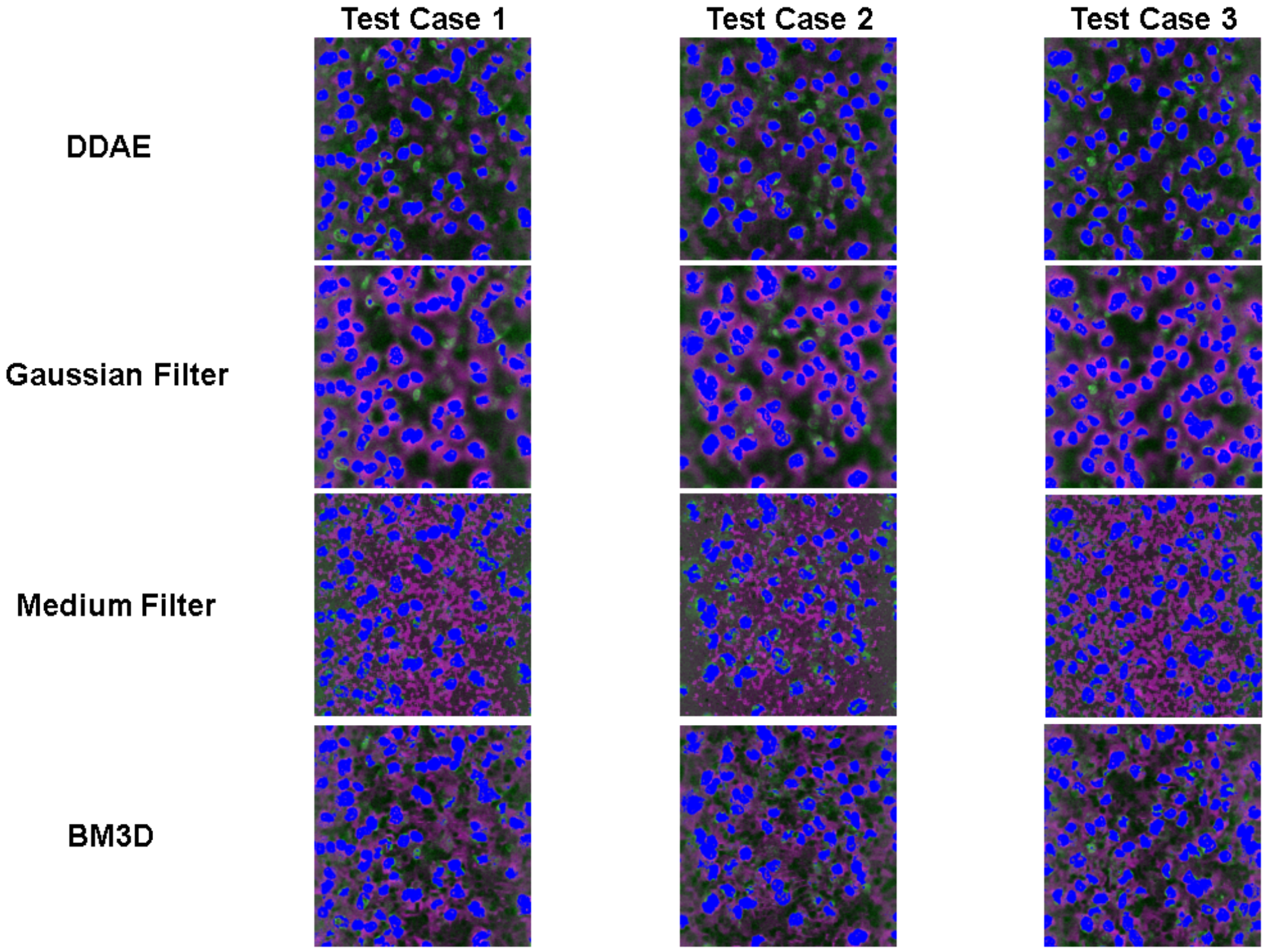

**Fig. 8.** Nuclear region analysis of 3PF images denoised with DDAE, Gaussian filter (σ=5), median filter (σ=10), and BM3D (σ=240) algorithms. Blue: true positive, Magenta: false positive, Green: false negative.

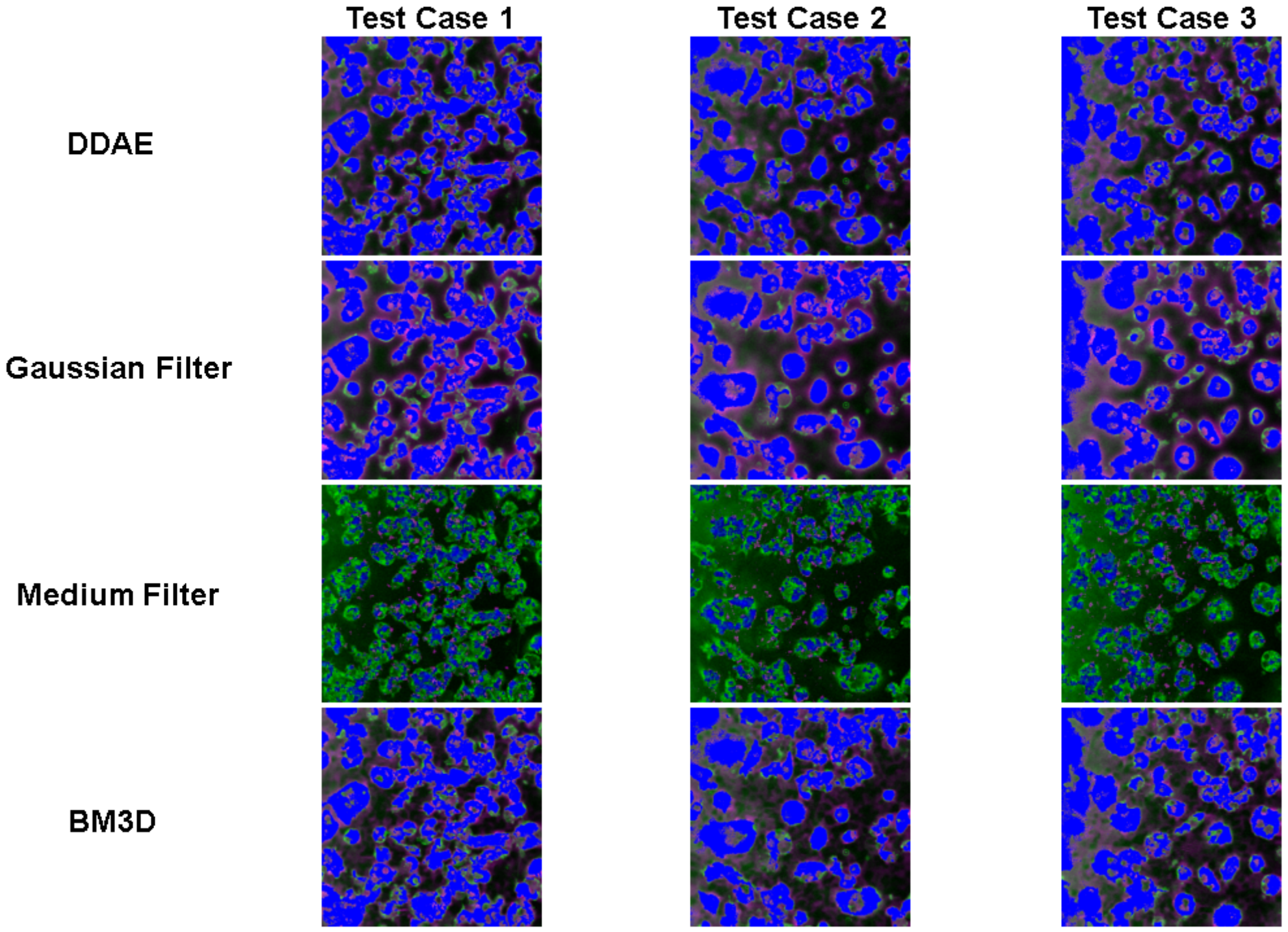

**Fig. 9.** Cellular region analysis of THG images denoised with DDAE, Gaussian filter (σ=5), median filter (σ=5), and BM3D (σ=180) algorithms. Blue: true positive, Magenta: false positive, Green: false negative.

**Table 2. The Precision, Recall, and F1- Score of the Denoising Results**[a]

| | 3PF Images | | | THG Images | | |
|---|---|---|---|---|---|---|
| **Filter** | **Precision** | **Recall** | **F1-score** | **Precision** | **Recall** | **F1-score** |
| DDAE | 0.70 | 0.77 | 0.734 | 0.76 | 0.88 | 0.821 |
| BM3D120 | 0.50 | 0.85 | 0.629 | 0.76 | 0.86 | 0.809 |
| BM3D140 | 0.48 | 0.87 | 0.617 | 0.76 | 0.87 | 0.814 |
| BM3D160 | 0.49 | 0.87 | 0.622 | 0.76 | 0.86 | 0.814 |
| BM3D180 | 0.47 | 0.88 | 0.610 | 0.76 | 0.87 | 0.818 |
| BM3D200 | 0.47 | 0.88 | 0.614 | 0.75 | 0.88 | 0.816 |
| BM3D220 | 0.48 | 0.87 | 0.621 | 0.76 | 0.87 | 0.816 |
| BM3D240 | 0.50 | 0.87 | 0.631 | 0.76 | 0.87 | 0.816 |
| Gaussian1 | 0.50 | 0.41 | 0.449 | 0.74 | 0.36 | 0.549 |
| Gaussian3 | 0.62 | 0.72 | 0.668 | 0.87 | 0.63 | 0.751 |
| Gaussian5 | 0.64 | 0.79 | 0.707 | 0.84 | 0.77 | 0.803 |
| Gaussian10 | 0.50 | 0.87 | 0.635 | 0.72 | 0.88 | 0.800 |
| Median1 | 0.40 | 0.23 | 0.288 | 0.47 | 0.38 | 0.426 |
| Median3 | 0.77 | 0.10 | 0.172 | 0.65 | 0.26 | 0.454 |
| Median5 | 0.94 | 0.07 | 0.132 | 0.70 | 0.21 | 0.453 |
| Median10 | 0.40 | 0.78 | 0.525 | 0.83 | 0.26 | 0.545 |

[a]Numbers in the filters column represent the sigma values of the algorithm.

## 4. DISCUSSION AND CONCLUSION

For low photon-budget multi-photon biomedical imaging, it's crucial to find a balanced denoising method to preserve stochastic resonance enhanced regions and retain cell boundary features for further segmentation. The de-speckle strategy of denoising will remove most of the noise boosted signals and corrode the region of cells. The low-pass spatial filtering strategy will sacrifice the resolution and expand the cellular regions. The non-local mean algorithm like BM3D average the patches with similar textures and achieve state-of-the-art performances. However, these methods involve either time-consuming optimization processes or manually chosen parameters which result in low computational efficiency when pursuing high performance. This problem becomes the analysis bottleneck of many high throughput 3D microscopies. Our results in compact DDAE model had uncovered its capability to realize both feature preservation and noise filtering. After several epochs of training, the validation loss can already be reduced greatly (Fig. S1 and S2). The optimization and parameter selection process can be accomplished in advance in the training of DDAE model, and the denoising process only cost a few seconds. Even at such high speed of denoising process, the DDAE model still outperformed other conventional methods on PSNR, SSIM, and the preservation of signal regions. This may be due to its nonlinear transformation characteristic of deep learning. The stochastic noise in the low photon-budget multiphoton imaging is not perturbative interference to signals. It can non-linearly boost the sub-threshold signals up above the detection limit. Hence, the methods such as Gaussian filters and BM3D may not be able to handle such a situation well in short computation time. Whereas, deep learning model could learn the non-linear features from the training dataset, effectively suppress the noise, and correctly preserve the cellular boundaries. In the present study, our promising results have validated the effectiveness of the DDAE system in the task of boundary-preserved image denoising. It is believed that such denoising performance of the proposed system can be further improved when combining new and more complex deep learning models.

Regarding the future works, the DDAE model has the potential to improve the image quality from various kind of image sources. We will implement our DDAE model on more different types of microscopy images for practical medical technology application. Besides, Lehtinen et al. showed that, without clean dataset, deep learning model can still perform good image restoration [35]. We will also try to train the neural network with cleaner images excited by higher laser power and detected at lower gain. A good denoising method can speed up the applications, such as high-throughput image segmentation, fast 3D morphodynamic analysis, and long-term cell tracking. We expect this denoising method can be organically integrated into those process of imaging analysis, help improve the processing efficiency of many high-throughput microscopies, and finally achieve on-line denoising on the hardware.

**Funding.** This work is sponsored by the Faculty of Health Sciences, University of Macau, the startup grants of the University of Macau, and supported by Science and Technology Development Fund (FDCT) of Macao SAR under grant numbers of 122/2016/A3 and 018/2017/A1. This work is also sponsored by Department of Internal Medicine (Division of Cardiology, Cardiovascular Center), National Taiwan University Hospital, and supported by Ministry of Science and Technology, R.O.C., under research plan of MOST 107-2314-B-002-262-MY2.

# Supplementary Information

### Supplementary Table 1. Denoising Performance of Filters on 3PF Images

| Gaussian Filter | case1 | case2 | case3 | average |
|---|---|---|---|---|
| *σ= 1* | | | | |
| PSNR | 21.21 | 21.6 | 21.18 | 21.33 |
| SSIM | 0.43 | 0.44 | 0.43 | 0.43 |
| *σ= 3* | | | | |
| PSNR | 25.69 | 25.8 | 25.28 | 25.59 |
| SSIM | 0.8 | 0.8 | 0.79 | 0.8 |
| *σ= 5* | | | | |
| PSNR | 25.69 | 25.67 | 25.16 | 25.51 |
| SSIM | 0.83 | 0.84 | 0.83 | 0.83 |
| *σ= 10* | | | | |
| PSNR | 24.83 | 24.68 | 24.22 | 24.58 |
| SSIM | 0.84 | 0.84 | 0.83 | 0.84 |

| Median Filter | case1 | case2 | case3 | average |
|---|---|---|---|---|
| *σ= 1* | | | | |
| PSNR | 12.37 | 12.76 | 12.4 | 12.51 |
| SSIM | 0.09 | 0.1 | 0.09 | 0.09 |
| *σ= 3* | | | | |
| PSNR | 18.53 | 19.04 | 18.42 | 18.66 |
| SSIM | 0.17 | 0.18 | 0.18 | 0.18 |
| *σ= 5* | | | | |
| PSNR | 18.61 | 19.1 | 18.45 | 18.72 |
| SSIM | 0.15 | 0.16 | 0.16 | 0.16 |
| *σ= 10* | | | | |
| PSNR | 18.45 | 18.87 | 18.26 | 18.53 |
| SSIM | 0.15 | 0.15 | 0.16 | 0.15 |

| BM3D | case1 | case2 | case3 | average |
|---|---|---|---|---|
| *σ= 120* | | | | |
| PSNR | 24.99 | 25.3 | 24.69 | 25 |
| SSIM | 0.71 | 0.72 | 0.71 | 0.71 |
| *σ= 140* | | | | |
| PSNR | 25.13 | 25.3 | 24.79 | 25.07 |
| SSIM | 0.71 | 0.73 | 0.71 | 0.72 |
| *σ= 160* | | | | |
| PSNR | 25.19 | 25.34 | 24.84 | 25.12 |
| SSIM | 0.72 | 0.73 | 0.72 | 0.72 |
| *σ= 180* | | | | |
| PSNR | 25.19 | 25.31 | 24.85 | 25.12 |
| SSIM | 0.71 | 0.72 | 0.72 | 0.72 |
| *σ= 200* | | | | |
| PSNR | 25.22 | 25.26 | 24.81 | 25.1 |
| SSIM | 0.71 | 0.72 | 0.71 | 0.71 |
| *σ= 220* | | | | |
| PSNR | 25.14 | 25.18 | 24.76 | 25.03 |
| SSIM | 0.71 | 0.72 | 0.71 | 0.71 |
| *σ= 240* | | | | |
| PSNR | 25.05 | 25.1 | 24.69 | 24.95 |
| SSIM | 0.71 | 0.71 | 0.71 | 0.71 |

| DDAE Model | case1 | case2 | case3 | average |
|---|---|---|---|---|
| PSNR | 26.69 | 26.89 | 26.21 | 26.6 |
| SSIM | 0.86 | 0.87 | 0.86 | 0.86 |

Yellow colors highlight the optimal PSNR for each testing case. Orange colors highlight the optimal average values of PSNR and its corresponding SSIM. For Gaussian filter, median filter, and BM3D, the optimal sigma value is chosen based on the average PSNR.

## Supplementary Table 2. Denoising Performance of Filters on THG Images

| Gaussian Filter | case1 | case2 | case3 | average |
|---|---|---|---|---|
| *σ= 1* | | | | |
| PSNR | 22.68 | 23.49 | 22.66 | 22.94 |
| SSIM | 0.5 | 0.54 | 0.49 | 0.51 |
| *σ= 3* | | | | |
| PSNR | 28.89 | 29.21 | 29.12 | 29.07 |
| SSIM | 0.84 | 0.85 | 0.84 | 0.84 |
| *σ= 5* | | | | |
| PSNR | 29.4 | 29.2 | 29.43 | 29.34 |
| SSIM | 0.85 | 0.87 | 0.87 | 0.86 |
| *σ= 10* | | | | |
| PSNR | 28.24 | 28.08 | 28.59 | 28.3 |
| SSIM | 0.82 | 0.86 | 0.86 | 0.85 |

| Median Filter | case1 | case2 | case3 | average |
|---|---|---|---|---|
| *σ= 1* | | | | |
| PSNR | 13.38 | 14.13 | 13.33 | 13.61 |
| SSIM | 0.17 | 0.21 | 0.15 | 0.18 |
| *σ= 3* | | | | |
| PSNR | 21.64 | 22.57 | 21.36 | 21.86 |
| SSIM | 0.27 | 0.32 | 0.24 | 0.28 |
| *σ= 5* | | | | |
| PSNR | 22.11 | 22.88 | 21.82 | 22.27 |
| SSIM | 0.23 | 0.28 | 0.2 | 0.24 |
| *σ= 10* | | | | |
| PSNR | 21.9 | 22.52 | 21.61 | 22.01 |
| SSIM | 0.21 | 0.26 | 0.18 | 0.22 |

| BM3D | case1 | case2 | case3 | average |
|---|---|---|---|---|
| *σ= 120* | | | | |
| PSNR | 29.25 | 29.62 | 29.42 | 29.43 |
| SSIM | 0.74 | 0.76 | 0.76 | 0.75 |
| *σ= 140* | | | | |
| PSNR | 29.33 | 29.64 | 29.5 | 29.49 |
| SSIM | 0.74 | 0.76 | 0.76 | 0.75 |
| *σ= 160* | | | | |
| PSNR | 29.28 | 29.58 | 29.55 | 29.47 |
| SIM | 0.74 | 0.76 | 0.76 | 0.75 |
| *σ= 180* | | | | |
| PSNR | 29.28 | 29.51 | 29.49 | 29.43 |
| SSIM | 0.74 | 0.75 | 0.76 | 0.75 |
| *σ= 200* | | | | |
| PSNR | 29.28 | 29.36 | 29.49 | 29.38 |
| SSIM | 0.73 | 0.74 | 0.76 | 0.74 |
| *σ= 220* | | | | |
| PSNR | 29.2 | 29.21 | 29.41 | 29.27 |
| SSIM | 0.73 | 0.74 | 0.75 | 0.74 |
| *σ= 240* | | | | |
| PSNR | 29.13 | 29.09 | 29.31 | 29.18 |
| SSIM | 0.72 | 0.73 | 0.75 | 0.73 |

| DDAE Model | case1 | case2 | case3 | average |
|---|---|---|---|---|
| PSNR | 29.7 | 30.16 | 29.91 | 29.92 |
| SSIM | 0.86 | 0.88 | 0.88 | 0.87 |

Yellow colors highlight the optimal PSNR in each testing case. Orange colors highlight the optimal average values of PSNR and its corresponding SSIM. For Gaussian filter, median filter, and BM3D, the optimal sigma value is chosen based on the average PSNR

# Supplementary Table 3. The Analysis of Boundary Preservation after Denoising

| 3PF | Test1 CNT$_{Ans}$=63171 | | Test2 CNT$_{Ans}$=57468 | | Test3 CNT$_{Ans}$=63621 | | Test1 | | | Test2 | | | Test3 | | | Average | | | Variance | | |
|---|---|---|---|---|---|---|---|---|---|---|---|---|---|---|---|---|---|---|---|---|---|
| Filter | CNT$_O$ | CNT$_\&$ | CNT$_O$ | CNT$_\&$ | CNT$_O$ | CNT$_\&$ | precision | recall | F1-score | precision | recall | F1-score | precision | recall | F1-score | precision | recall | F1-score | precision | recall | F1-score |
| BM3D120 | 108887 | 53800 | 109223 | 50098 | 97102 | 53152 | 0.49409 | 0.85166 | 0.62537 | 0.45868 | 0.87175 | 0.60109 | 0.54738 | 0.83545 | 0.66141 | 0.50005 | 0.85295 | 0.62929 | 0.20% | 0.03% | 0.09% |
| BM3D140 | 117717 | 54663 | 114822 | 50776 | 102050 | 54450 | 0.46436 | 0.86532 | 0.60439 | 0.44221 | 0.88355 | 0.58942 | 0.53356 | 0.85585 | 0.65733 | 0.48005 | 0.86824 | 0.61705 | 0.23% | 0.02% | 0.13% |
| BM3D160 | 105512 | 53533 | 116523 | 51014 | 106530 | 54911 | 0.50736 | 0.84743 | 0.63472 | 0.4378 | 0.88769 | 0.5864 | 0.51545 | 0.8631 | 0.64544 | 0.48687 | 0.86607 | 0.62218 | 0.18% | 0.04% | 0.10% |
| BM3D180 | 123242 | 55563 | 115865 | 50984 | 106988 | 55171 | 0.45084 | 0.87956 | 0.59613 | 0.44003 | 0.88717 | 0.58828 | 0.51567 | 0.86718 | 0.64675 | 0.46885 | 0.87797 | 0.61039 | 0.17% | 0.01% | 0.10% |
| BM3D200 | 122087 | 55620 | 114783 | 51051 | 106051 | 55012 | 0.45558 | 0.88047 | 0.60046 | 0.44476 | 0.88834 | 0.59275 | 0.51873 | 0.86468 | 0.64845 | 0.47302 | 0.87783 | 0.61389 | 0.16% | 0.01% | 0.09% |
| BM3D220 | 119832 | 55571 | 110896 | 50719 | 103915 | 54838 | 0.46374 | 0.87969 | 0.60732 | 0.45736 | 0.88256 | 0.60249 | 0.52772 | 0.86195 | 0.65464 | 0.48294 | 0.87473 | 0.62149 | 0.15% | 0.01% | 0.08% |
| BM3D240 | 116231 | 55209 | 106601 | 50372 | 100729 | 54521 | 0.47499 | 0.87396 | 0.61548 | 0.47253 | 0.87652 | 0.61403 | 0.54126 | 0.85697 | 0.66347 | 0.49626 | 0.86915 | 0.631 | 0.15% | 0.01% | 0.08% |
| DDAE | **70631** | **48858** | **62780** | **44046** | **69709** | **49331** | **0.69174** | **0.77342** | **0.7303** | **0.70159** | **0.76644** | **0.73259** | **0.70767** | **0.77539** | **0.73998** | **0.70033** | **0.77175** | **0.73429** | **0.01%** | **0.00%** | **0.00%** |
| Gaussian10 | 107181 | 53873 | 100232 | 49868 | 111174 | 56015 | 0.50264 | 0.85281 | 0.63249 | 0.49753 | 0.86775 | 0.63244 | 0.50385 | 0.88045 | 0.64092 | 0.50134 | 0.867 | 0.63528 | 0.00% | 0.02% | 0.00% |
| Gaussian1 | 47152 | 23814 | 59336 | 27471 | 47082 | 24674 | 0.50505 | 0.37698 | 0.43171 | 0.46297 | 0.47802 | 0.47038 | 0.52406 | 0.38783 | 0.44577 | 0.49736 | 0.41428 | 0.44929 | 0.10% | 0.31% | 0.04% |
| Gaussian3 | 77490 | 46357 | 72642 | 42893 | 64420 | 43970 | 0.59823 | 0.73383 | 0.65913 | 0.59047 | 0.74638 | 0.65933 | 0.68255 | 0.69112 | 0.68681 | 0.62375 | 0.72378 | 0.66843 | 0.26% | 0.08% | 0.03% |
| Gaussian5 | 84475 | 51188 | 71462 | 45298 | 71652 | 49044 | 0.60595 | 0.81031 | 0.69339 | 0.63388 | 0.78823 | 0.70268 | 0.68447 | 0.77088 | 0.72511 | 0.64143 | 0.78981 | 0.70706 | 0.16% | 0.04% | 0.03% |
| Median10 | 133879 | 50824 | 81738 | 39389 | 159234 | 54975 | 0.37963 | 0.80455 | 0.51585 | 0.48189 | 0.68541 | 0.56591 | 0.34525 | 0.8641 | 0.49337 | 0.40226 | 0.78469 | 0.52504 | 0.51% | 0.83% | 0.14% |
| Median1 | 27327 | 11626 | 49404 | 16832 | 31994 | 13892 | 0.42544 | 0.18404 | 0.25693 | 0.3407 | 0.29289 | 0.31499 | 0.43421 | 0.21836 | 0.29058 | 0.40012 | 0.23176 | 0.2875 | 0.27% | 0.31% | 0.08% |
| Median3 | 3955 | 3348 | 13997 | 9316 | 6829 | 5494 | 0.84652 | 0.053 | 0.09975 | 0.66557 | 0.16211 | 0.26072 | 0.80451 | 0.08636 | 0.15597 | 0.7722 | 0.10049 | 0.17215 | 0.90% | 0.31% | 0.67% |
| Median5 | 717 | 717 | 10456 | 8987 | 3713 | 3629 | 1 | 0.01135 | 0.02245 | 0.85951 | 0.15638 | 0.26462 | 0.97738 | 0.05704 | 0.10779 | 0.94563 | 0.07492 | 0.13162 | 0.57% | 0.55% | 1.51% |

| THG | Test1 CNT$_{Ans}$=111440 | | Test2 CNT$_{Ans}$=94364 | | Test3 CNT$_{Ans}$=101969 | | Test1 | | | Test2 | | | Test3 | | | Average | | | Variance | | |
|---|---|---|---|---|---|---|---|---|---|---|---|---|---|---|---|---|---|---|---|---|---|
| Filter | CNT$_O$ | CNT$_\&$ | CNT$_O$ | CNT$_\&$ | CNT$_O$ | CNT$_\&$ | precision | recall | F1-score | precision | recall | F1-score | precision | recall | F1-score | precision | recall | F1-score | precision | recall | F1-score |
| BM3D120 | 126433 | 95768 | 110216 | 82644 | 111736 | 87051 | 0.75746 | 0.85937 | 0.8052 | 0.74984 | 0.8758 | 0.80794 | 0.77908 | 0.8537 | 0.81468 | 0.76212 | 0.86296 | 0.80927 | 0.02% | 0.01% | 0.00% |
| BM3D140 | 128732 | 97052 | 111997 | 83483 | 114154 | 87914 | 0.75391 | 0.87089 | 0.80819 | 0.7454 | 0.88469 | 0.8091 | 0.77014 | 0.86216 | 0.81356 | 0.75648 | 0.87258 | 0.81453 | 0.02% | 0.01% | 0.00% |
| BM3D160 | 119645 | 92985 | 111639 | 83587 | 115472 | 88677 | 0.77717 | 0.8344 | 0.80477 | 0.74873 | 0.88579 | 0.81151 | 0.76795 | 0.86965 | 0.81564 | 0.76462 | 0.86328 | 0.81395 | 0.02% | 0.07% | 0.00% |
| BM3D180 | 128021 | 97135 | 109971 | 83235 | 114331 | 88222 | 0.75874 | 0.87163 | 0.81128 | 0.75688 | 0.88206 | 0.81469 | 0.77164 | 0.86518 | 0.81574 | 0.76242 | 0.87296 | 0.81769 | 0.01% | 0.01% | 0.00% |
| BM3D200 | 126902 | 96674 | 117369 | 85423 | 113281 | 87795 | 0.7618 | 0.8675 | 0.81122 | 0.72782 | 0.90525 | 0.80689 | 0.77502 | 0.861 | 0.81575 | 0.75488 | 0.87792 | 0.8164 | 0.06% | 0.06% | 0.00% |
| BM3D220 | 124715 | 95692 | 114643 | 84716 | 111277 | 86990 | 0.76729 | 0.85869 | 0.81042 | 0.73895 | 0.89776 | 0.81065 | 0.78174 | 0.8531 | 0.81587 | 0.76266 | 0.86985 | 0.81625 | 0.05% | 0.06% | 0.00% |
| BM3D240 | 122265 | 94462 | 111146 | 83638 | 117024 | 89301 | 0.7726 | 0.84765 | 0.80839 | 0.75251 | 0.88633 | 0.81396 | 0.7631 | 0.87577 | 0.81556 | 0.76274 | 0.86992 | 0.81633 | 0.01% | 0.04% | 0.00% |
| DDAE | **128648** | **98203** | **111842** | **84121** | **117476** | **89424** | **0.76335** | **0.88122** | **0.81806** | **0.75214** | **0.89145** | **0.81589** | **0.76121** | **0.87697** | **0.815** | **0.7589** | **0.88321** | **0.82106** | **0.00%** | **0.01%** | **0.00%** |
| Gaussian10 | 133282 | 95923 | 121391 | 85103 | 120869 | 89541 | 0.7197 | 0.86076 | 0.78393 | 0.70107 | 0.90186 | 0.78889 | 0.74081 | 0.87812 | 0.80364 | 0.72053 | 0.88025 | 0.80039 | 0.04% | 0.04% | 0.01% |
| Gaussian1 | 41632 | 32917 | 54607 | 38530 | 53654 | 38471 | 0.79067 | 0.29538 | 0.43009 | 0.70559 | 0.40831 | 0.51728 | 0.71702 | 0.37728 | 0.49441 | 0.73776 | 0.36032 | 0.54904 | 0.21% | 0.34% | 0.20% |
| Gaussian3 | 76181 | 66488 | 72319 | 62863 | 75045 | 64923 | 0.87276 | 0.59663 | 0.70875 | 0.86925 | 0.66618 | 0.75428 | 0.86512 | 0.63669 | 0.73354 | 0.86904 | 0.63317 | 0.7511 | 0.00% | 0.12% | 0.05% |
| Gaussian5 | 99653 | 82808 | 86558 | 73484 | 93817 | 78971 | 0.83096 | 0.74307 | 0.78456 | 0.84896 | 0.77873 | 0.81233 | 0.84176 | 0.77446 | 0.80671 | 0.84056 | 0.76542 | 0.80299 | 0.01% | 0.04% | 0.02% |
| Median10 | 34163 | 28377 | 29170 | 24364 | 32940 | 27179 | 0.83062 | 0.25463 | 0.38978 | 0.83522 | 0.25819 | 0.39444 | 0.8251 | 0.26654 | 0.40292 | 0.83031 | 0.25979 | 0.54505 | 0.00% | 0.00% | 0.00% |
| Median1 | 84781 | 42974 | 80668 | 35534 | 84663 | 39280 | 0.50688 | 0.38562 | 0.43802 | 0.4405 | 0.37656 | 0.40603 | 0.46396 | 0.38522 | 0.42094 | 0.47045 | 0.38247 | 0.42646 | 0.11% | 0.00% | 0.03% |
| Median3 | 40479 | 28664 | 43050 | 24515 | 39752 | 26525 | 0.70812 | 0.25721 | 0.37736 | 0.56945 | 0.25979 | 0.3568 | 0.66726 | 0.26013 | 0.37433 | 0.64828 | 0.25904 | 0.45366 | 0.51% | 0.00% | 0.01% |
| Median5 | 31936 | 23031 | 29802 | 20225 | 31246 | 21521 | 0.72116 | 0.20667 | 0.32127 | 0.67865 | 0.21433 | 0.32577 | 0.68875 | 0.21105 | 0.3231 | 0.69618 | 0.21068 | 0.45343 | 0.05% | 0.00% | 0.00% |

CNT$_O$: **count$_O$**; CNT$_\&$: **count$_{AND}$**; CNT$_{Ans}$: **count$_{Ans}$.** Yellow: precision higher than DDAE's. Blue-green: recall > 80%. Pink: F1-score > 0.65 in 3PF and > 0.8 in THG images.

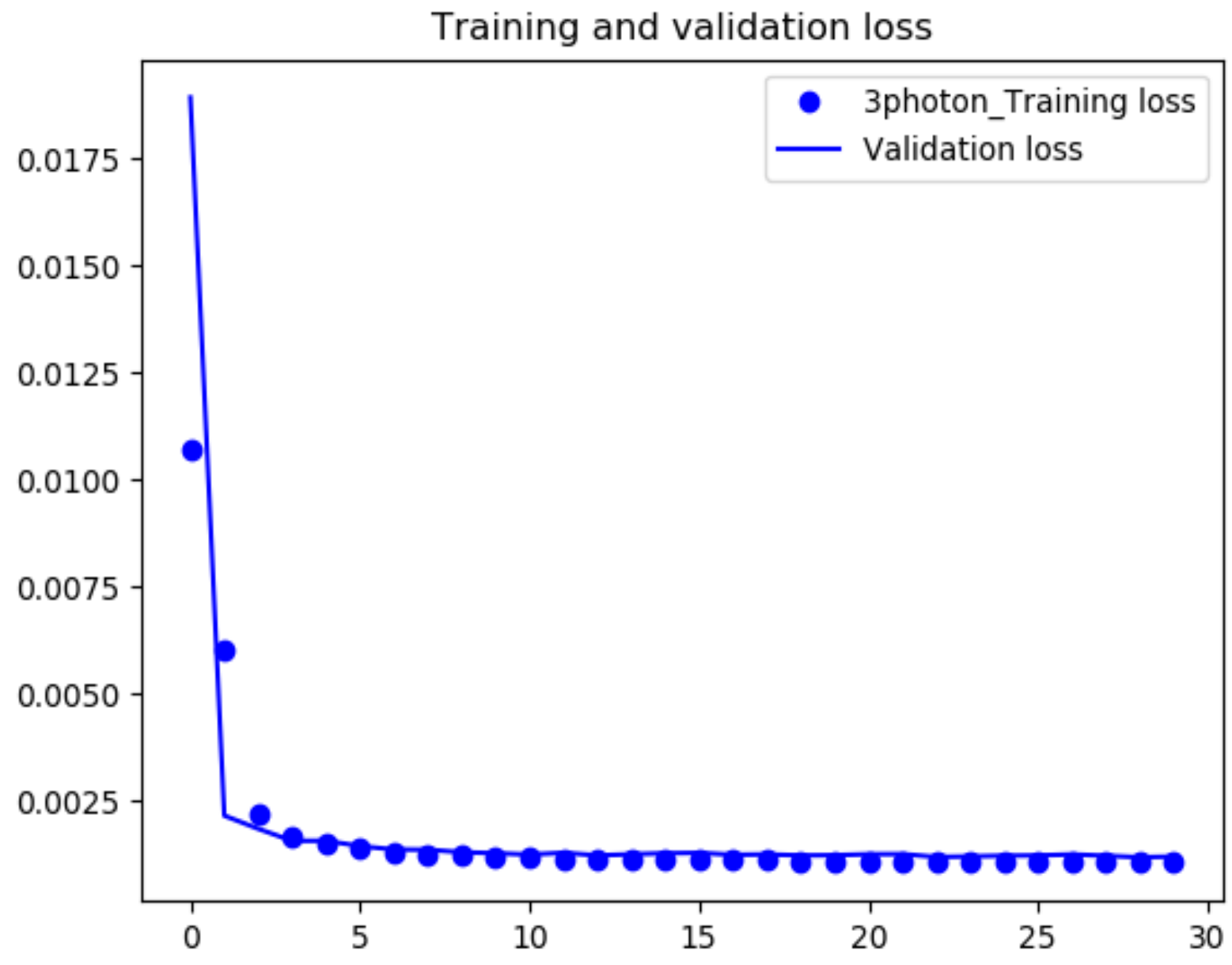

**Fig. S1**. Training and validation loss across different epoch iteration during the DDAE training of 3PF images.

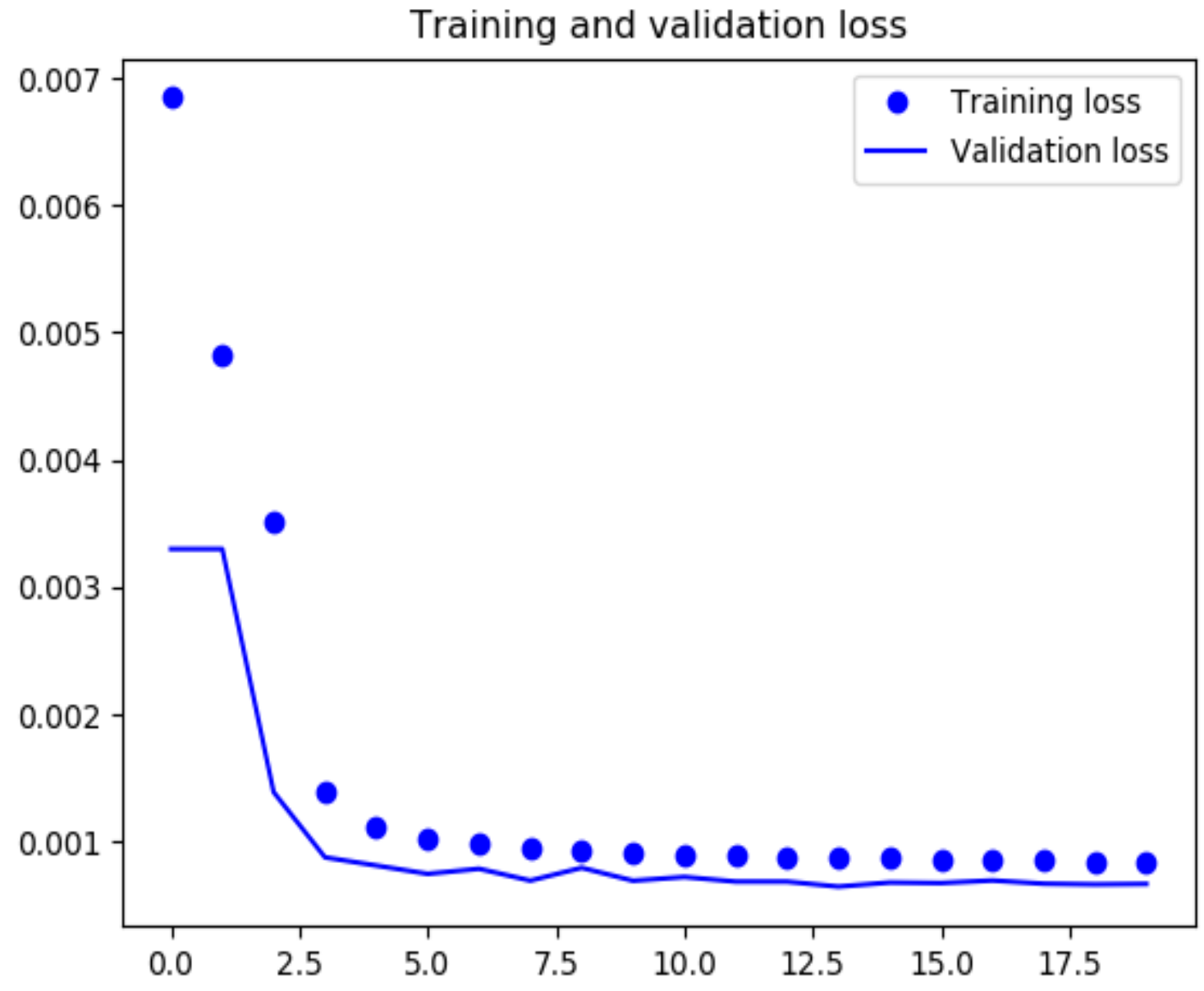

**Fig. S2.** Training and validation loss across different epoch iteration during the DDAE training of THG images.